%% file: TaxoKG-Fusion.tex
\documentclass[sigconf]{acmart}

\copyrightyear{2023}
\acmYear{2023}
\setcopyright{rightsretained}
\acmConference[SIGIR '23]{Proceedings of the 46th International ACM SIGIR Conference on Research and Development in Information Retrieval}{July 23--27, 2023}{Taipei, Taiwan}
\acmBooktitle{Proceedings of the 46th International ACM SIGIR Conference on Research and Development in Information Retrieval (SIGIR '23), July 23--27, 2023, Taipei, Taiwan}\acmDOI{10.1145/3539618.3591997}
\acmISBN{978-1-4503-9408-6/23/07}

\usepackage{caption}
\usepackage{subcaption}
\usepackage{multirow}
\usepackage{graphicx}
\usepackage{hyperref}
\usepackage{booktabs}
\usepackage{subcaption}
\usepackage{commath}
\usepackage{xcolor}
\usepackage{enumerate}
\usepackage{xspace}
\usepackage{makecell}
\usepackage{enumitem}
\usepackage{csquotes}

\usepackage{pgfplots}
\usepackage{pgfplotstable}
\usetikzlibrary{pgfplots.groupplots}
\usepgfplotslibrary{colorbrewer}
\usepackage{tikz}
\usetikzlibrary{positioning}
\pgfplotsset{compat=newest}

\usepackage{bbm}
\usepackage{amsmath}

\usepackage{amsthm}
\theoremstyle{definition}
\newtheorem{definition}{Definition}[section]

\definecolor{myblue1}{HTML}{2a6a99}
\definecolor{myblue2}{HTML}{7494b5}
\definecolor{myblue3}{HTML}{b1c0d1}
\definecolor{myorange1}{HTML}{d88546}
\definecolor{myorange2}{HTML}{e5a77d}
\definecolor{myorange3}{HTML}{edcab4}
\definecolor{mypurple}{HTML}{cfd5eb}

\newcommand{\ours}{HiPrompt\xspace}
\newcommand{\benchmark}{KG-\textsc{Hi}-BKF\xspace}
\newcommand{\header}[1]{\noindent \textbf{#1}}
\newcommand{\eg}[0]{\textit{e.g.}\xspace}

\begin{document}

\title[HiPrompt: Few-Shot Biomedical Knowledge Fusion via Hierarchy-Oriented Prompting]{HiPrompt: Few-Shot Biomedical Knowledge Fusion via Hierarchy-Oriented Prompting}


\author{Jiaying Lu}
\affiliation{%
  \institution{Emory University, USA}
}
\email{jiaying.lu@emory.edu}
\author{Jiaming Shen}
\affiliation{%
  \institution{Google Research, USA}
}
\email{jmshen@google.com}
\author{Bo Xiong}
\affiliation{%
  \institution{University of Stuttgart, Germany}
}
\email{bo.xiong@ipvs.uni-stuttgart.de}
\author{Wenjing Ma}
\affiliation{%
  \institution{Emory University, USA}
}
\email{wenjing.ma@emory.edu}
\author{Steffen Staab}
\affiliation{%
  \institution{University of Stuttgart, Germany \\ University of Southampton, UK}
}
\email{steffen.staab@ipvs.uni-stuttgart.de}
\author{Carl Yang}
\affiliation{%
  \institution{Emory University, USA}
}
\email{j.carlyang@emory.edu}

\input{tex/abstract}




\maketitle

\input{tex/introduction}
\input{tex/approach}
\input{tex/experiments}
\input{tex/conclusion} 
\input{tex/acknowledgement}

\bibliographystyle{ACM-Reference-Format}
\bibliography{TaxoKG-Fusion-ref}


\end{document}

%% file: tex/abstract.tex
\begin{abstract}
Medical decision-making processes can be enhanced by comprehensive biomedical knowledge bases, which require fusing knowledge graphs constructed from different sources via a uniform index system. The index system often organizes biomedical terms in a hierarchy to provide the aligned entities with fine-grained granularity. To address the challenge of scarce supervision in the biomedical knowledge fusion (BKF) task, researchers have proposed various unsupervised methods. However, these methods heavily rely on ad-hoc lexical and structural matching algorithms, which fail to capture the rich semantics conveyed by biomedical entities and terms. Recently, neural embedding models have proved effective in semantic-rich tasks, but they rely on sufficient labeled data to be adequately trained. To bridge the gap between the scarce-labeled BKF and neural embedding models, we propose \ours, a supervision-efficient knowledge fusion framework that elicits the few-shot reasoning ability of large language models through hierarchy-oriented prompts. Empirical results on the collected \benchmark benchmark datasets demonstrate the effectiveness of \ours. 
\end{abstract}

%% file: tex/introduction.tex

\section{Introduction}
\label{sec:introduction}

\begin{figure}[htp!]
    \centering
    \vspace*{-1.0em}
    \setlength{\abovecaptionskip}{0cm}
    \includegraphics[width=0.98\linewidth]{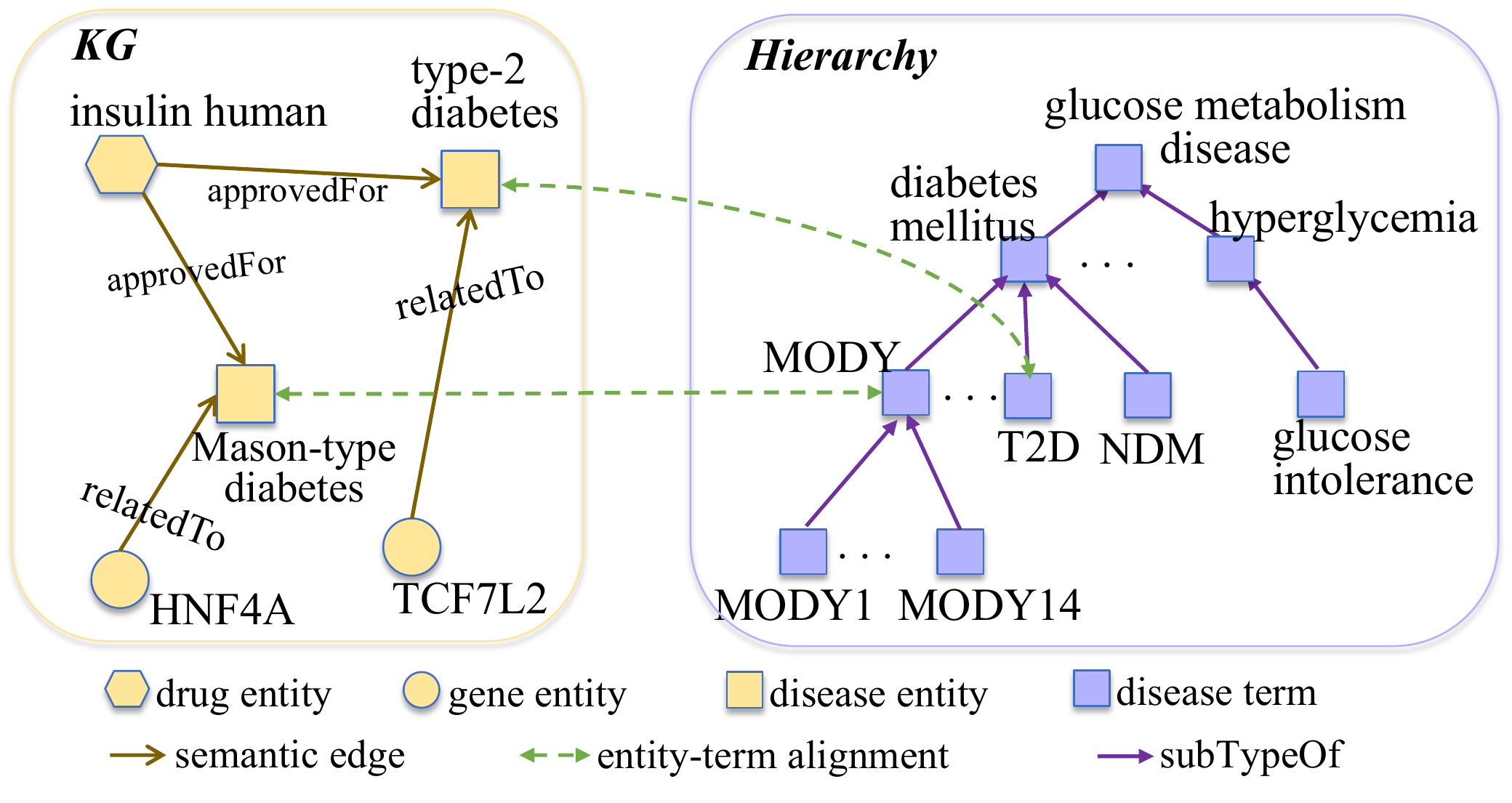}
    \caption{A toy example of BKF to find entity-term alignment between KG and hierarchy. \emph{Left}: A KG containing biomedical entities. \emph{right}: A hierarchy containing biomedical terms.} 
    \label{fig:toy_example}
    \vspace*{-1.0em}
\end{figure}

\begin{figure*}[ht!]
    \centering
    \setlength{\abovecaptionskip}{0cm}
    \includegraphics[width=0.98\linewidth]{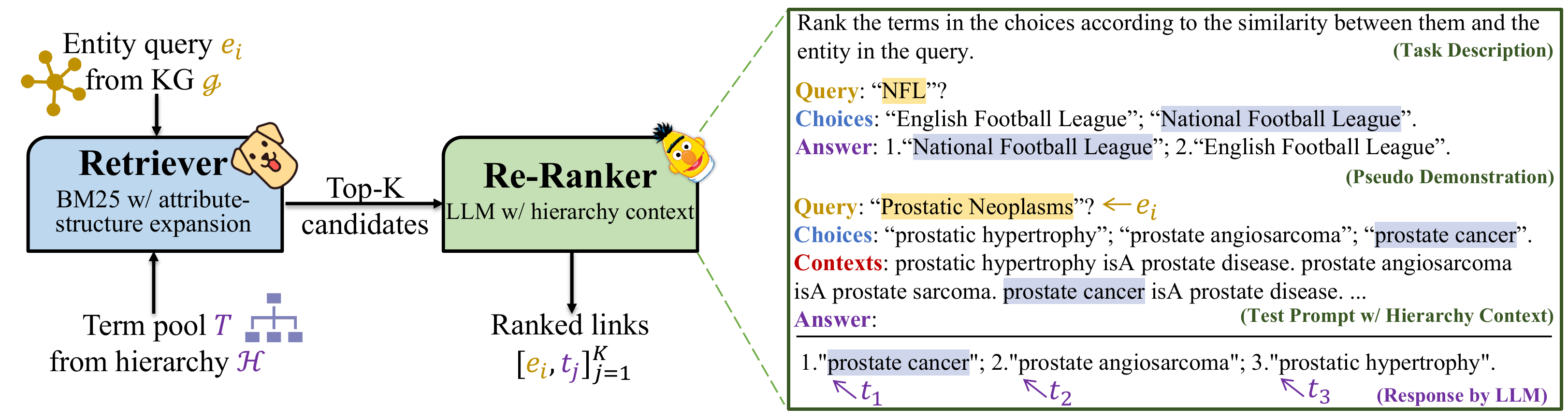}
    \caption{Overview of our \ours framework, with a zoom-in on the LLM-based re-ranker.}
    \label{fig:framework_two_col}
    \vspace*{-1.2em}
\end{figure*}

In the biomedical field, there exists a lot of knowledge acquired from clinical practice guidelines, medical records, and publications, accumulated from different research laboratories and healthcare institutions~\cite{Sigdel2019CloudBasedPM,su2021iBKH,chandak2022preMedKG}. Recently, knowledge graphs (KGs) have emerged as a compelling technique to efficiently represent, organize, and distribute knowledge. 
A biomedical KG stores the properties of biomedical entities and their relations.
Researchers' constant endeavors in manually curating biomedical KGs have led to the existence of many domain-specific and application-oriented KGs.
However, these well-annotated biomedical KGs are scattered in various data formats, which hinders their off-the-shelf usability.

Fusing KGs from multiple sources into an accurate and comprehensive knowledge base can greatly support clinical decision-making~\cite{himmelstein2017Hetionet,santos2020CKG}. 
A common practice is to align entities of KGs with standard hierarchical index systems (\textit{i.e.} \emph{biomedical hierarchies})~\cite{schriml2022DO,jiang2023cellTaxo,wishart2018drugbank,bodenreider2004umls}. The hierarchy allows entities to be aligned and analyzed more precisely with fine-grained granularity, which is beneficial to many downstream tasks~\cite{shen2021taxoclass,tsatsaronis2015overview,wang2022hierarchical,shen2022automated,ma23cellcano}. 
Moreover, the biomedical hierarchy is well maintained with periodic upgrades to incorporate newly emerging biomedical terms, thus enabling scalable integration with multiple KGs.
In this work, we study the biomedical knowledge fusion (\emph{BKF}) problem that aims to align entities from biomedical KGs into terms from the biomedical hierarchy. Figure~\ref{fig:toy_example} gives a toy example of the BKF task. The BKF task is challenging due to the following characteristics. 
First, inconsistent naming vocabularies are used in different resources, as they are developed independently by different groups of specialists. Second, unlike the existing KG entity alignment problem~\cite{sun2020benchmarkingEA,DBLP:conf/www/XuSX022} that contains many labeled entity-entity pairs as training samples, biomedical knowledge integration is supervision-scarce. Third, the topology of a KG and a hierarchy are very different, where the KG is a general graph, while the hierarchy is a directed acyclic graph.

\header{Existing research.} Pioneer studies on BKF mainly rely on the biomedical thesaurus to normalize words and match lexical to establish alignment between KGs and the hierarchy~\cite{himmelstein2017Hetionet,Ren2017LifeiNetAS,su2021iBKH,santos2020CKG}. Later, researchers explore combing first-order logic~\cite{jimenez2011logmap}, probabilistic alignment~\cite{suchanek2011paris}, or non-literal string comparisons~\cite{faria2013aAML} with lexical matching for unsupervised BKF. However, these methods fail to capture the rich semantics conveyed in entities and terms (\eg, synonyms, definitions, types), which are essential to handle the inconsistent naming conventions from multi-sources. 
Another line of work leverage neural embedding models~\cite{chen2017MTransE,lu22:hakeGCN,sun2020benchmarkingEA,xiong2022faithful,lu23gtd2g} to represent entities as dense vectors using semantic attributes, structural properties, and alignment supervisions. These models perform better than unsupervised models when sufficient training samples are available. However, the scarcity of supervision in the BKF problem leads to the underfitting of these data-eager neural models.
Moreover, none of the existing methods explicitly leverages the hierarchical structure of terms in the biomedical hierarchy.

\header{Present work.} To address above challenges, we present \textbf{\ours}, a few-shot BKF framework via \textbf{Hi}erarchy-Oriented \textbf{Promp}ting. 
\ours employs a large language model (LLM) to generatively propose terms from the hierarchy to be aligned with entities from the KG.
The key insight is that LLMs~\cite{brown2020GPT3,zhang2022opt,GPTJT,dong23closed} can be rapidly adapted to an unseen task via the gradient-free ``prompt-based learning''~\cite{wang2022towards,singhal2022MedPaLM}, thus removing the dependencies on the task-specific supervision. \ours applies prompt-based learning with a curated task description for the BKF task and a tiny number of demonstrations generated from the few-shot samples. This mimics the procedure of how humans accomplish a new task by learning from previous experiences and generalizing them to a new context. Moreover, we add the hierarchical context to the prompts to further improve the performance of \ours. To evaluate the performance of our proposed \ours, we create \emph{\benchmark}, a new benchmark for BKF with two datasets collected from two biomedical KGs~\cite{zhu2022SDKG,brown2017repoDB} and one disease hierarchy~\cite{schriml2022DO} with manual verification. 
Empirical results demonstrate the effectiveness of our \ours framework, which largely outperforms both conventional unsupervised lexical matching models and neural semantic embedding models.

%% file: tex/approach.tex
\section{Biomedical Knowledge Fusion} 

\subsection{Problem Definition} \label{sec:problem-definition}
BKF aims at aligning existing specialized biomedical KGs into a uniform biomedical index system that can be represented by a hierarchy. 
We define the biomedical KG and hierarchy as follows: A biomedical KG is a multi-relation graph $\mathcal{G}=(E,R,RT)$, where $E,R,RT$ are a set of various types of entities, a set of relation names, and $RT\in E\times R \times E$ is the set of relational triples, respectively. A biomedical hierarchy is a directed acyclic graph (DAG) $\mathcal{H}=(T,TP)$, where $T$ is a set of terms, and $TP\in T\times T$ is a set of hypernym-hyponymy term pairs, respectively.  
The topology differences between KG and hierarchy distinguish our BKF task from other related tasks (\eg, entity alignment, KG integration).
Moreover, both entities $E$ and terms $T$ contain rich associated semantic attributes (\eg, definition, synonyms). 
Finally, we define our task as follows:

\vspace{-\topsep}
\begin{definition}[biomedical knowledge fusion]
Given a biomedical KG $\mathcal{G}$, a biomedical hierarchy $\mathcal{H}$, a set of pre-aligned entity-term pairs $[e_a,t_a]_{a=1}^M$, and a set of unaligned entities $[e_1,e_2,\cdots, e_N] \in \mathcal{G}$.  
The goal is to link each unaligned entity to the hierarchy $LK=\{(e_i, t_j) | e_i \in \mathcal{G}, t_j \in \mathcal{H}\}$ such that $t_j$ is the most specific term in the hierarchy for entity $e_i$ in KG. 
In our work, we focus on the few-shot settings where the sample size $M$ is very small to reflect the scarcity of labeled data that is ubiquitous in the biomedical field. 
\end{definition}
\vspace{-\baselineskip}

\begin{table*}[htbp!]
\scalebox{0.90}{
\begin{tabular}{ll|cccccc|cccccc}
\toprule
\multirow{2}{*}{Setting} & \multirow{2}{*}{Model} & \multicolumn{6}{c}{SDKG-DzHi} & \multicolumn{6}{c}{repoDB-DzHi} \\
 & & Hits@1& Hits@3& {\small nDCG@1}& {\small nDCG@3}& WuP & MRR & Hits@1& Hits@3 & {\small nDCG@1}& {\small nDCG@3}& WuP & MRR \\
\midrule
\multirow{8}{*}{Zero-shot} & Edit Dist & 65.51& 70.39& 68.08&  50.82& 85.53& 68.69&  68.69& 71.37&71.71& 54.15&  85.21&  70.71\\
& BM25 & 73.07& 87.40& 77.56& 63.01& 91.97 & 81.06& 59.38& 74.75& 70.33& 64.51& 90.71& 68.84\\
& LogMap & 75.75& 79.06& 76.97& 54.82& 85.06 & 77.38 & 86.60& 87.73& 87.38& 60.79& 91.68 &  87.09\\
& PARIS & 22.68& 22.68& 23.15& 16.13& 43.85& 22.68& 6.35& 6.35& 6.42& 4.44&  32.28 &  6.35\\
& AML & OOM& OOM& OOM & OOM & OOM & OOM& 78.00& 78.56& 78.67&  54.90&  86.02& 78.26\\
& SapBERT & 69.61 & 87.24&76.38& 63.86& 93.78& 78.97& 75.04& 90.69& 81.24& 73.51& 94.25 & 83.61\\
& SelfKG  &  57.95& 69.45&  58.98& 47.29&  74.25&  64.70& 72.78& 81.10& 75.95& 63.78&  88.41&  77.71\\
& \ours  &  \textbf{90.79}& \textbf{93.08}& \textbf{91.57}& \textbf{77.00}& \textbf{96.74}& \textbf{92.13}& \textbf{88.01}& \textbf{91.26}& \textbf{90.70}& \textbf{82.85}& \textbf{97.06}& \textbf{90.64}\\
\midrule
\multirow{3}{*}{One-shot} & SapBERT &  69.56& 87.22&76.34&  63.84& 93.29& 78.93 &  75.00& 90.68&81.21& 73.51 & 94.13& 83.59\\
& MTransE &  0.0& 0.16& 0.0& 0.05& 35.09& 0.16& 0.0&  0.28&  0.14&  0.27& 28.89& 0.37\\
& \ours  & \textbf{92.11}& \textbf{95.11}& \textbf{93.53}& \textbf{77.63}& \textbf{97.25}& \textbf{93.91}&  \textbf{88.28}& \textbf{91.53}& \textbf{90.61}& \textbf{81.31}& \textbf{96.39} & \textbf{90.28}\\
\bottomrule
\end{tabular}
}
\caption{Main experiment results (in percentages).}
\vspace*{-2.0em}
\label{tab:main_exp}
\end{table*}


\subsection{Technical Details of \ours} \label{ssec:approach}

Figure~\ref{fig:framework_two_col} shows the overall architecture of our proposed \ours framework. To tackle the BKF task with limited training samples, our key insight is to utilize LLMs via hierarchy-oriented prompting. However, LLMs can not accommodate very lengthy input prompts (\eg, GPT-3 only supports up to 4096 tokens) that contain all candidate terms along with their hierarchy contexts. A feasible workaround is to exhaustively examine each candidate term given the query entity, but the inference cost would be dramatic~\cite{patterson2021carbon}.  
Therefore, we propose to use the \emph{retrieve and re-rank}~\cite{wang2011cascade,matsubara2020reranking,glass2022re2g} approach to resolve the above challenges.

\smallskip
\header{Retrieval Module.} The retriever provides an efficient solution for coarse-grained candidate filtering, thus reducing the overall inference cost of \ours. Given one entity query $e_i$ from the KG $\mathcal{G}$ and all candidate terms $T$ from the hierarchy $\mathcal{H}$, the retriever produces a coarsely ranked candidate list $(t'_1,t'_2,\cdots,t'_K)$, to avoid unnecessary computations for the LLM-based re-ranker. 
\ours framework is flexible so that any unsupervised ranking function (\eg, TF-IDF~\cite{salton1988term}, LDA~\cite{blei2003latent}) can be used to generate the ranked list.
In practice, we choose the unsupervised BM25~\cite{robertson2009BM25} as the ranking function. Since entities and concepts have rich attributive and structural information, we further utilize these two types of information to expand~\cite{billerbeck2005expansion} query entities and candidate terms.



 
\smallskip
\header{Re-Ranking Module.} Given the query entity $e_i$ and the coarsely ranked candidate list $(t'_1,t'_2,\cdots,t'_K)$, we request the LLM to re-rank the list to $(t_1,t_2,\cdots,t_K)$ where $t_1$ is the most specific term of $e_i$ via the gradient-free prompt-based learning.
Figure~\ref{fig:framework_two_col} provides an example of the input prompt and the response of the re-ranker. The input prompt is composed of (1) curated textual \emph{\textbf{task description}}, (2) illustrative \emph{\textbf{demonstration}} from few-show samples, and (3) the \emph{\textbf{test prompt}} constructed from the query entity and the coarsely ranked list.
The LLM-based re-ranker essentially tackles the BKF task by estimating the conditional probability: $P_{LLM}(w_1,w_2,\dots,w_n | prompt)$, where $(w_1,\dots,w_n)$ is the output word sequence with variable lengths. The desired re-ranked list can be converted from the output sequence by a simple mapping function $(t_1,t_2,\cdots,t_K)=f(w_1,w_2,\dots,w_n)$. 

For the template of demonstration, we use the query entity to form the question string ``Query: \{$e_i$\}'', the coarse candidate list to form the choice string ``Choices: \{$t'_1$; $t'_2$; $\dots$ $t'_K$\}'', and the ground truth to form the answer string ``Answer: \{$t_1$; $t_2$; $\dots$, $t_K$\}''. While there is no such ground truth sample in the zero-shot setting, we propose the \emph{\textbf{pseudo demonstration}} technique which adopts out-of-domain entity-term pairs to showcase what is the perspective format. Both real and pseudo demonstrations are essential to generate output sequences in the consistent format~\cite{schick2021exploiting,kojima2022large}. For the test prompt, we use the same template of the demonstration, while leaving the answer string as ``Answer:'' for LLM to predict what comes next. To further elicit LLMs with hierarchical constraints and dependencies of candidate terms, we propose the novel \emph{\textbf{test prompt with hierarchy context}} where hypernyms of each candidate term are included in the context string. More specifically, we traverse the biomedical hierarchy $\mathcal{T}$ to locate the hypernym terms $t'_{i,p_1},\cdots,t'_{i,p_j}$ of a candidate term $t'_i$. Therefore, the context string is formed as ``Contexts: \{$t'_1$ isA $t'_{1,p}$; \dots; $t'_K$ isA $t'_{K,p}$\}''.

%% file: tex/experiments.tex
\section{Experiments}
\label{sec:experiments}

\header{Benchmark Datasets.} 
We use the following data sources to create our \benchmark benchmark\footnote{\benchmark benchmark is available at \url{https://doi.org/10.6084/m9.figshare.21950282}.}: (1) SDKG~\cite{zhu2022SDKG}: a disease-centric KG that covers five cancers and six non-cancer diseases. (2) repoDB~\cite{brown2017repoDB}: we adopt their original triples, and generate entity attributes by querying DrugBank~\cite{wishart2018drugbank} and UMLS Metathesaurus~\cite{bodenreider2004umls}. (3) DzHi~\cite{schriml2022DO}: a hierarchy derived from the widely used Disease Ontology~\cite{schriml2022DO} which has a depth of $13$.
We first use the mapping existing in the resources themselves, which leads to many-to-many linkages between two KBs. We further manually verify the correctness of the many-to-many linkages and curate the datasets to the correct stage.
Table~\ref{tab:taxoKG-link} shows the statistics of the created benchmark. As can be seen, the linkages follow the one-to-one assumption~\cite{sun2020benchmarkingEA}, and the scale of labeled entity-term pairs is very small.

\begin{table}[ht!]
\centering
\vspace*{-0.5em}
\begin{tabular}{c|c|cc|c}
\toprule
Dataset                        & Source & \#Disease & \#Entities & \#Links \\
\midrule
\multirow{2}{*}{{\small SDKG-DzHi}}   & SDKG   &  841         & 19,416     &  635          \\
                & DzHi & 11,159    & 11,159     &  635          \\
\hline
\multirow{2}{*}{{\small repoDB-DzHi}} & repoDB & 2,074          & 3,646      &  709              \\
    & DzHi & 11,159    & 11,159     &  709      \\
\bottomrule
\end{tabular}
\caption{Statistics of the \benchmark benchmark.} 
\label{tab:taxoKG-link}
\vspace*{-2.5em}
\end{table}

\header{Compared Models.} We compare \ours to the following two sets of baselines: (a) \emph{Non-neural conventional models}: (a.1) \textbf{Edit Dist} ~\cite{ristad1998editDist} that quantifies the distance between entities and terms by the edit distance of their names. (a.2) \textbf{BM25}~\cite{robertson2009BM25} that ranks a set of documents based on the query tokens appearing in each document. 
(a.3) \textbf{LogMap}~\cite{jimenez2011logmap} that matches entities and terms via logical constraints and semantical features. (a.4) \textbf{PARIS}~\cite{suchanek2011paris} that provides a off-the-shelf fusion tool empowered by a parameter tuning-free probabilistic model. (a.5) \textbf{AML}~\cite{faria2013aAML} that is based on non-literal string comparison algorithms.
is a probabilistic matching system based on probability estimates. 
(b) \emph{Neural embedding models}: (b.1) \textbf{SapBERT}~\cite{liu2021SapBERT} that learns to self-align synonymous biomedical entities through a Transformer. 
(b.2) \textbf{MTransE}~\cite{chen2017MTransE} that extends the translational KG embedding method TransE~\cite{Bordes2013TransE} to multi-language system entity alignment by axis calibration and linear transformations. (b.3) \textbf{SelfKG}~\cite{liu2022selfkg} that designs a self-negative sampling strategy to push sampled negative pairs far away from each other when no labeled positive pairs are available.

\header{Quantitative evaluations.} We mainly focus on zero-shot and one-shot settings, and utilize the remaining labeled samples as the test set to report quantitative results.
Several \emph{strict} and \emph{lenient} evaluation metrics are used.
For strict metrics that appreciate only the exact correct prediction, we adopt \textbf{Hits@k} and mean reciprocal rank (\textbf{MRR}). For lenient metrics that also reward near-hits, we adopt \textbf{nDCG@k} with exponential decay~\cite{balog2012hierarchical} and hierarchy-based term relatedness score \textbf{WuP} ~\cite{wup1994verbs}.
All compared baselines are executed with their recommended hyperparameters.
For all non-neural conventional models, we only report the zero-shot results as they are unsupervised methods. 
For neural embedding methods, we report the zero-shot results utilizing released model weights (SapBERT) or conducting self-supervised training (SelfKG), while reporting the one-shot results by fine-tuning these models (SapBERT, MTransE) on the one demonstrative training sample.
For our \ours, we use GPT-3~\cite{brown2020GPT3} as the LLM for re-ranker and set its temperature hyperparameters as $0$ to lower the completion randomness. 
Using a single prompt template is sufficient since initial exploration shows that various templates do not have a significant impact on model performance.
We exclude the use of automatic prompt generation techniques~\cite{shin2020autoprompt,zhang2023autoCoT} due to the limited availability of training data.

\header{Main Results.} Table~\ref{tab:main_exp} shows the quantitative results for zero-shot and one-shot settings. \ours largely outperforms all other methods in all evaluation metrics under both settings, which demonstrates the effectiveness of the proposed hierarchy-oriented prompting. Under the zero-shot setting, the non-neural unsupervised baseline LogMap achieves the second-best performance. 
All examined models can successfully generate predictions except AML throws out-of-memory (OOM) errors on the SDKG-DzHi dataset. PARIS performs worst in the zero-shot setting because it can not predict aligned terms for each query entity. Instead, PARIS produces the alignment based on its own ad-hoc threshold. MTransE performs worst in the one-shot setting since it is underfitting using just one training sample.
Comparing the same models (SapBERT, HiPrompt) between zero-shot and one-shot settings, we observe the performance differences are negligible, thus indicating that effectively eliciting the adaptive reasoning ability is one of the key factors to tackling supervision-scarce BKF problem.

\begin{table}[ht!]
\vspace*{-1.0em}
\resizebox{\linewidth}{!}{%
\begin{tabular}{c|ccc|ccc}
\toprule
\multirow{2}{*}{Expan.} & \multicolumn{3}{c}{\textit{SDKG-DzHi}} & \multicolumn{3}{c}{\textit{repoDB-DzHi}}  \\
& Hits@5  & Hits@10  & Hits@20 & Hits@5  & Hits@10  & Hits@20\\
\midrule
Name & 88.66& 89.61& 90.55& 85.05& 88.72& 90.27\\ 
+Atr.& 94.96& 96.85& 98.11& 89.00& 92.52& 95.20\\ 
+Str. & 90.08& 90.71& 91.81& 88.15& 90.27& 92.24\\ 
+Atr.+Str.& \textbf{96.85}& \textbf{97.64}& \textbf{98.74}& \textbf{91.11}& \textbf{93.65}& \textbf{95.63}\\ 
\bottomrule
\end{tabular}%
}
\caption{Retriever with various expansion strategies.}
\vspace*{-2.5em}
\label{tab:retrieval_module}
\end{table}

\begin{table}[ht!]
\vspace*{-0.5em}
\resizebox{\linewidth}{!}{%
\begin{tabular}{c|cccccc}
\toprule
\multirow{2}{*}{LLMs} & \multicolumn{3}{c}{SDKG-DzTaxo} & \multicolumn{3}{c}{repoDB-DzTaxo} \\
& Hits@1 & Hits@3 & MRR  & Hits@1 & Hits@3 & MRR  \\
\midrule
   & \multicolumn{6}{c}{\textit{One-shot} (prompt w/o Hi. Context)} \\
GPT-3 & \textbf{91.80}& \textbf{94.32}& \textbf{93.45}&  \textbf{87.85}& \textbf{91.24}& \textbf{89.92}\\
GPT-JT &  75.08& 86.44 & 81.80 &  58.33& 69.77 & 66.42\\
OPT-6.7B &  68.93&  80.44& 76.38& 60.73& 73.59& 69.33\\
\midrule
   & \multicolumn{6}{c}{\textit{One-shot} (prompt w/ Hi. Context)} \\
GPT-3  &  \textbf{92.11}& \textbf{95.11}& \textbf{93.91}& \textbf{88.28}& \textbf{91.53}& \textbf{90.28}\\
GPT-JT &  80.76& 93.69 &  87.45&  69.07&  82.91& 77.24\\
OPT-6.7B & 72.40& 84.86& 79.64& 63.70& 77.68& 72.41\\
\bottomrule
\end{tabular}%
}
\caption{Re-ranker with various LLMs and prompts.}
\label{tab:LMs_exp}
\vspace*{-2.5em}
\end{table}

\header{Ablation Studies.} We further conduct ablation studies to evaluate the impact of our hierarchy-oriented techniques. 
Table~\ref{tab:retrieval_module} compares the different expansion strategies for \ours's retrieval module. As can be seen, if expanding the KG entities and hierarchy terms with both attributive and structural features (``\emph{+Atr.+Str.}'' variant), the retriever can achieve the best Hits@K performance. 
Table~\ref{tab:LMs_exp} compares different LLMs and different prompts for \ours's re-ranking module. Among the examined LLMs, GPT-3 with 175 billion parameters surpasses GPT-JT~\cite{GPTJT} with 6B parameters and OPT-6.7B ~\cite{zhang2022opt} with 6.7B parameters due to its large parameter space. When adding the proposed hierarchy context to the name-only prompts, every LLM achieves better performance on all metrics, thus demonstrating the importance of explicit hierarchy-oriented information. We also observe that improvements for GPT-JT and OPT-6.7B are more significant than GPT-3, since GPT-3 may already have such hierarchical information encoded.


\begin{figure}[ht!]
    \centering
    \setlength{\abovecaptionskip}{0cm}
    \vspace*{-0.5em}
    \includegraphics[width=\linewidth]{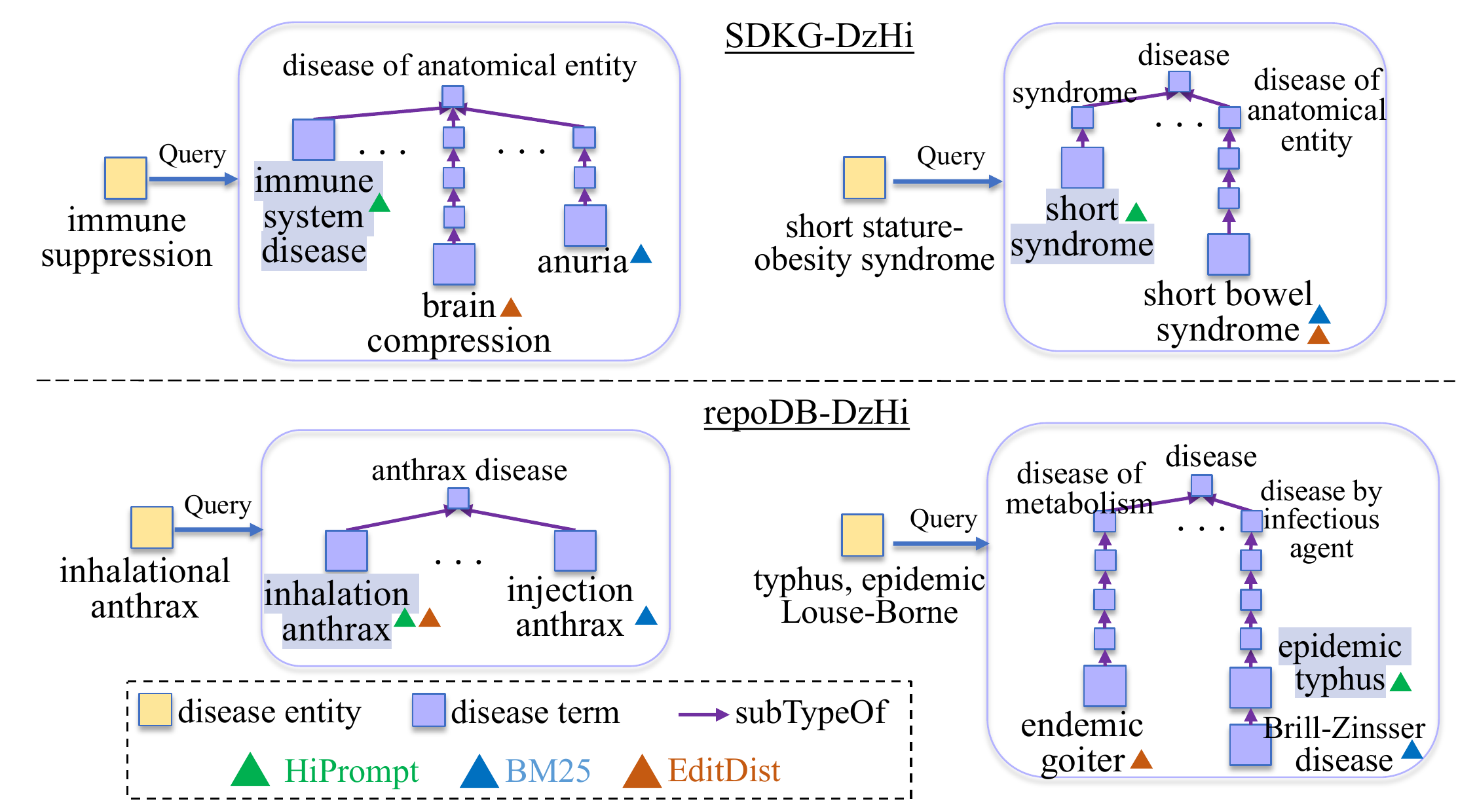}
    \caption{Case Studies on unlabeled data. Terms highlighted \colorbox{mypurple}{in violet} denote the correct alignments for query entities.}
    \label{fig:case_study}
    \vspace*{-1.0em}
\end{figure}

\header{Case Studies.} Figure~\ref{fig:case_study} shows the fusion results from BM25, EditDist, and \ours. In general, \ours can find the most specific terms in the hierarchy for the query entities, by satisfying the semantic similarities and hierarchical constraints simultaneously. For instance, \ours recognizes that ``\emph{immune system disease}'' is the most appropriate for the query ``\emph{immune suppression}'', rather than its hypernym ``\emph{disease of anatomical entity}'' that is too general, or hyponyms such as ``\emph{immune system cancer}'' or ``\emph{allergic disease}'' that are too specific.
On the other hand, EditDist only considers lexical matching, thereby ignoring the different naming conventions of the same biomedical concepts. BM25 also mainly relies on lexical matching, but it incorporates the names, definitions, and synonyms of biomedical terms during the matching, resulting in better performance in handling various names. However, BM25 ignores the hierarchical information, which leads to the inappropriate granularity of aligned terms (\eg, the term ``\emph{epidemic typhus}'' is too broad for the query entity ``typhus, epidemic Louse-Borne'').

%% file: tex/conclusion.tex
\section{Conclusions}
\label{sec:conclusion}

This paper studies how to automatically fuse KGs into a standard hierarchical index system with scarce labeled data. Our novel framework, \ours, uses hierarchy-oriented prompts to elicit the few-shot reasoning ability of large language models and is designed to be supervision-efficient. Performance comparison on the newly collected \benchmark benchmark with two datasets demonstrates the effectiveness of \ours. Interesting future directions for BKF include: (1) exploring an automatic way to generate hierarchy-aware prompts to further reduce manual intervention; (2) expanding the scope of biomedical knowledge fusion to allow the hierarchy to dynamically grow with the aligned entities. 

%% file: tex/acknowledgement.tex
\section*{Acknowledgement}
This research is supported by the internal fund and GPU servers provided by the Computer Science Department of Emory University.
Bo Xiong was supported by the International Max Planck Research School for Intelligent Systems (IMPRS-IS) and was funded by the European Union’s Horizon 2020 research and innovation programme under the Marie Skłodowska-Curie grant agreement No: 860801 and Deutsche Forschungsgemeinschaft (DFG, German Research Foundation) under Germanys' Excellence Strategy—EXC 2075-390740016 (SimTech).